\pdfoutput=1
\documentclass[a4paper,british,aps,pre]{revtex4}
\usepackage[T1]{fontenc}
\usepackage[latin1]{inputenc}
\usepackage{float}
\usepackage{graphicx}

\makeatletter

\usepackage{babel}
\makeatother
\begin{document}

\title{Topological Classification of Clusters in Condensed Phases}

\author{Stephen R. Williams}

\affiliation{Research School of Chemistry, The Australian National University,
Canberra, ACT 0200, Australia.}

\email{swilliams@rsc.anu.edu.au}

\date{\today}

\begin{abstract}
A methodology is developed to identify clusters in a bulk phase which
are topologically equivalent to certain reference clusters. The selected
reference clusters are the Morse clusters with 13 or less atoms and
the 13 atom clusters found in an FCC and an HCP crystal phase, consisting
of an atom and its 12 nearest neighbours. The first step in using
the method requires the bond network of the bulk phase to be identified.
The bonds may be identified in terms of the distance between the atom
pairs or by using a modified Voronoi decomposition, introduced here.
We then search for all the 3, 4 and 5 membered shortest path rings
and use these to identify some basic clusters. These basic clusters
are then combined to identify the above mentioned Morse and crystal
clusters. The larger Morse clusters ($N>13$) may be identified in
terms of the basic clusters and the smaller Morse clusters. This work
has important applications in studying the energy landscape of a bulk
phase. As an example, results from a limited preliminary study on
the hard sphere fluid phase are presented.
\end{abstract}
\maketitle

\section{Introduction}

We consider dense bulk atomic phases. An understanding of the energy
landscape of such bulk phases has important applications to vitrification
\cite{Debenedetti-Nat-01}, crystallisation \cite{Frenkel-NM-2006},
protein folding \cite{Wales-Book} and colloidal dispersions \cite{Royall-tbp-07}.
Due to the large number of degrees of freedom in such systems, a direct
approach to these problems is often extremely difficult, even with
the use of modern computers. Recently the study of small clusters,
of isolated particles, has been the subject of important progress
\cite{Alonso-Book,Wales-Book}. To date, attempts to bring this progress
to bear on bulk phases have been limited. Here we pursue this end.
Obviously the identification of clusters, similar to those which in
isolation are in an energy minimum, in bulk phases is an important
part of the energy landscape problem. Due to their efficient packing
\cite{Sloane-DCG-1995,Manoharan-sci-2003} and low energy, such clusters
should be thermodynamically favoured in supercooled liquids at low
temperatures and high pressures \cite{Frenkel-NM-2006}. We develop
the methodology necessary to identify clusters in a bulk phase, which
are topologically similar to given reference clusters. The reference
clusters are isolated configurations located at minima in the interparticle
potential energy landscape. For the Morse clusters \cite{DOYE1995,morse-clusters}
(that we focus on here) the Morse potential is used, which is qualitatively
similar to a common Lennard-Jones potential with the additional feature
that the range of its potential well may be varied by a free parameter.
Many of the reference clusters we focus on are similar to those obtained
for configurations of spheres, which minimise the second moment of
the mass distribution \cite{Manoharan-sci-2003}. We use the notation
found in references \cite{DOYE1995,morse-clusters} to denote the
various Morse clusters. 

The identification of clusters in a bulk phase at finite temperature
is not an easy task. The clusters may undergo significant perturbations
due to the thermal motions and the stresses induced by the rest of
the system. Exactly how large a perturbation may be, before we decide
that a cluster is no longer similar to the given reference cluster,
is a decision which involves arbitrary choices. The goal is to maximise
the permitted perturbation while still excluding configurations that
are obviously incorrect. The means by which we pursue this inexact
goal will become apparent as the paper proceeds. Regardless of these
difficulties, there are many methodologies that could be developed
to give excellent results at low enough temperatures or high enough
densities. Under such conditions the larger perturbations will be
so heavily suppressed that they become insignificant. It is such a
methodology that we introduce here.

\section{Shortest Path Rings}

Our method makes use of 3, 4 and 5 membered shortest path (SP) rings
\cite{FRANZBLAU1991}. A shortest path ring is defined in terms of
the shortest distance, $D$, between a pair of atoms. If two atoms
are bonded together this distance is $D=1$ and if two atoms are not
bonded but they are both bonded to a common third atom then $D=2$,
etc. An $n$ membered ring has $n$ atoms which are bonded together
to form a ring. To understand what a shortest path ring is we consider
$n$ atoms, taken from a bulk phase, which form an $n$ membered ring.
In isolation the $n$ membered ring forms a $n$ membered graph. The
bulk phase, consisting of $N$ atoms, also forms a graph. If the distance
between each of the atoms in the graph $n$ is the same as the distance
between the same atoms in the graph $N$ and each atom in graph $n$
has exactly two bonds the ring is a shortest path ring. We denote
an $n$ membered shortest path ring as SP\emph{n.}

\section{Determining The Bond Network}

The simplest way to determine the bond network (what was referred
to as a graph in section II) is to label any pair of atoms which are
closer together than some distance $r_{b}$ as bonded. This method
can be useful when the pair potential between the two particles features
an attractive well which is deeper than several $k_{B}T$ such as
in a colloidal gel phase. Such an approach is not so useful for a
dense supercritcal fluid phase which is commonly used as a model to
study vitrification and crystallisation.

To handle this type of system it is necessary to use a different method
to determine the bond network. The standard Voronoi decomposition
\cite{Bernal-DOTFS-1967,Finney-PRSL-1970} is not suitable for the
identification of SP4 rings. For this reason we need to modify the
Voronoi method. First let us consider a method by which a bond network
may be established from a Voronoi tessellation. The network consists
of $N$ atoms where each atom is labelled by the index $i$. In the
Voronoi tessellation each atom is assigned the volume in its immediate
neighbourhood consisting of the space which is closer to it than to
any of the other atoms. The surface of this volume is composed of
flat polygons shared between two atoms. If two atoms share a surface
and the line which connects their centres intersects this surface
we define them as bonded, otherwise they are not. Let us assume that
all the bonded atoms are separated by some length less than $r_{c}$
which may be made as large as necessary. We start with the atom $i=1$
and iterate up. We construct the set of atoms, $S_{i}$, by finding
all the atoms that are within distance $r_{c}$ of atom $i$. The
position of a given atom is represented by $\mathbf{r}_{i}$ and the
atoms in $S_{i}$ are ordered starting with the atom closest to $i$.
We iterate through the atoms in $S_{i}$ using the index $k$ (from
closest to furthest from $i$), and construct the vector $\mathbf{u}=\mathbf{r}_{k}-\mathbf{r}_{i}$.
We eliminate from $S_{i}$ all the atoms, that are further from $i$
than $k$ is, and that are on the other side of the plane (relative
to $\mathbf{r}_{i}$) that is perpendicular to $\mathbf{u}$ and contains
the point $\mathbf{r}_{k}$. We then continue eliminating atoms from
$S_{i}$ by iterating up through $k$. The atoms which remain in $S_{i}$
are bonded to $i$ and, of course $i$ is bonded to the atoms which
remain in $S_{i}$. We can then iterate $i$ and continue until all
the bonds have been established. In mathematical form, atoms $i$
and $j$ are not bonded if they do not satisfy the inequality,\begin{equation}
\mathbf{r}_{i}\cdot\mathbf{r}_{k}+\mathbf{r}_{j}\cdot\mathbf{r}_{k}<\mathbf{r}_{k}\cdot\mathbf{r}_{k}+\mathbf{r}_{i}\cdot\mathbf{r}_{j},\label{eq:st-voronoi}\end{equation}
for any of the other atoms $k$. 

The problem with using the standard Voronoi method to identify SP4
rings can be appreciated by considering a ring formed by placing atoms
exactly on the corners of a perfect square. For such an arrangement
to form an SP4 ring the atoms on opposite corners must not be bonded.
Using Eq. \ref{eq:st-voronoi} we see that if atoms $i$ and $j$
are on opposite corners, with atom $k$ being one of the remaining
two atoms, we have $\mathbf{r}_{i}\cdot\mathbf{r}_{k}+\mathbf{r}_{j}\cdot\mathbf{r}_{k}=\mathbf{r}_{k}\cdot\mathbf{r}_{k}+\mathbf{r}_{i}\cdot\mathbf{r}_{j}$
and thus there will be many instances where we fail to identify SP4
rings which are required. 

To overcome this we modify the previous algorithm. As a first step
we consider what happens if we eliminate atoms that are past the plane
which is perpendicular to $\mathbf{u}$ and contains the point $f_{c}\mathbf{r}_{k}$,
where $f_{c}$ is some arbitrary parameter. This allows us to move
the plane closer to the $ith$ atom and eliminate bonds which are
impeding the identification of the required SP4 rings. Thus we eliminate
atoms which fail to satisfy the following inequality,\begin{equation}
\mathbf{r}_{i}\cdot\mathbf{r}_{i}-\mathbf{r}_{i}\cdot\mathbf{r}_{j}-\mathbf{r}_{i}\cdot\mathbf{r}_{k}+\mathbf{r}_{j}\cdot\mathbf{r}_{k}<f_{c}\left(\mathbf{r}_{i}\cdot\mathbf{r}_{i}+\mathbf{r}_{k}\cdot\mathbf{r}_{k}-2\mathbf{r}_{i}\cdot\mathbf{r}_{k}\right).\label{eq:mod-vor-right}\end{equation}
It is easy to see that Eq. \ref{eq:mod-vor-right} reduces to Eq.
\ref{eq:st-voronoi} when $f_{c}=1$. As it stands this procedure
has a serious flaw. If we swap the indices $i$ \& $j$, in Eq. \ref{eq:mod-vor-right},
when $f_{c}\neq1$ we may get a different answer: i.e. atom $i$ may
be bonded to atom $j$ while atom $j$ is not bonded to atom $i$.
The inequality, after we swap the indices, is\begin{equation}
\mathbf{r}_{j}\cdot\mathbf{r}_{j}-\mathbf{r}_{i}\cdot\mathbf{r}_{j}+\mathbf{r}_{i}\cdot\mathbf{r}_{k}-\mathbf{r}_{j}\cdot\mathbf{r}_{k}<f_{c}\left(\mathbf{r}_{j}\cdot\mathbf{r}_{j}+\mathbf{r}_{k}\cdot\mathbf{r}_{k}-2\mathbf{r}_{j}\cdot\mathbf{r}_{k}\right).\label{eq:mod-vor-left}\end{equation}
We can add these two inequalities together to obtain,

\begin{equation}
\mathbf{r}_{i}\cdot\mathbf{r}_{i}+\mathbf{r}_{j}\cdot\mathbf{r}_{j}-2\mathbf{r}_{i}\cdot\mathbf{r}_{j}<f_{c}\left(\mathbf{r}_{i}\cdot\mathbf{r}_{i}+\mathbf{r}_{j}\cdot\mathbf{r}_{j}+2\mathbf{r}_{k}\cdot\mathbf{r}_{k}-2\mathbf{r}_{i}\cdot\mathbf{r}_{k}-2\mathbf{r}_{j}\cdot\mathbf{r}_{k}\right),\label{eq:mod-voronoi}\end{equation}
which remains symmetric upon exchanging indices $i$ and $j$. We
may use this criteria Eq. \ref{eq:mod-voronoi} in place of criteria
Eq. \ref{eq:st-voronoi} and tune the value of $f_{c}$ so that we
identify the SP4 rings appropriately. Here we set the arbitrary parameter
$f_{c}$ to some value less than unity $f_{c}<1$ in an attempt to
optimise the effective identification of the clusters which is our
goal. If we set $f_{c}=1$ we regain the standard Voronoi method. 

\begin{figure}
\includegraphics[scale=0.6]{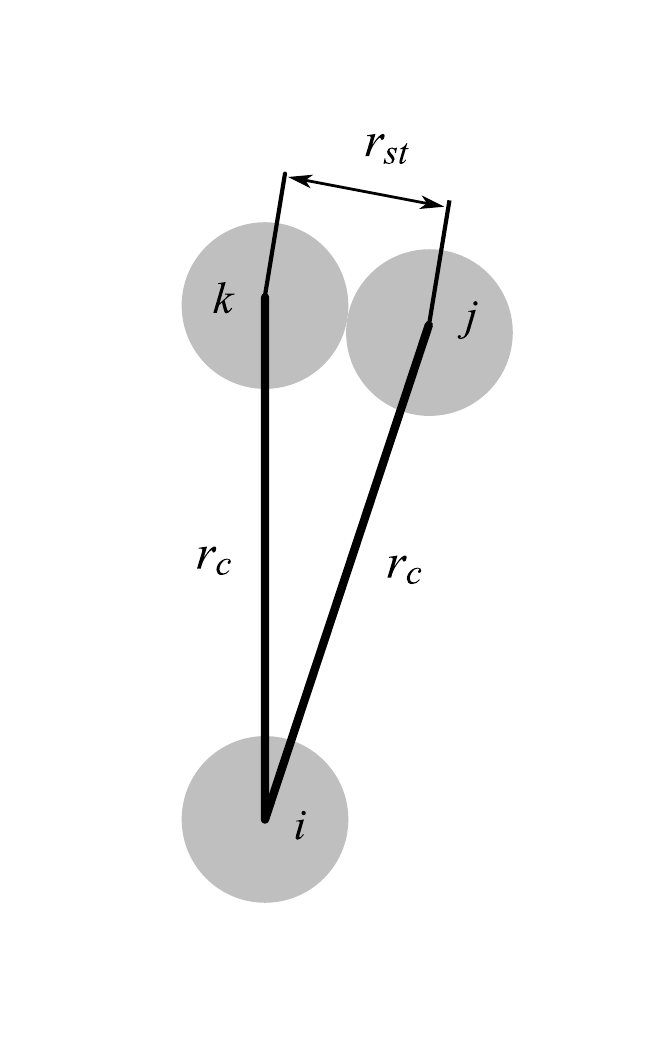}

\caption{Atoms $k$ and $j$ are as close together as their steric interaction
will permit, $r_{st}$, and are both the same distance from atom $i$,
namely $r_{c}$. \label{fig-1}}
\end{figure}

It is necessary to be careful not to make $f_{c}$ too small. Consider
the three atom system shown in Fig \ref{fig-1} where we might expect
that atoms $j$ and $k$ are bonded to atom $i$. %
 However if $f_{c}$ is lower than the following value, 

\begin{eqnarray}
f_{c} & = & \frac{1}{1+2\sin^{2}(\theta)}\nonumber \\
\theta & = & 2\,\sin^{-1}\left(\frac{r_{st}}{2r_{c}}\right),\label{eq:fc}\end{eqnarray}
given in terms of the distance, $r_{c}$, of atom $j$ or $k$ from
atom $i$, and the distance, $r_{st}$, between the two atoms $j$
\& $k$, (in Fig. \ref{fig-1} depicted as the closest distance the
steric interaction will allow) then atoms $j\,\&\, k$ will not be
identified as bonded to $i$. As our algorithm assumes atoms can only
be eliminated from $S$ by considering atoms $k$ which are closer
to atom $i$ than $j$ we must choose a value for $r_{c}$ which is
less than that obtained from Eq. \ref{eq:fc}. 

\begin{figure}
\includegraphics[scale=0.6]{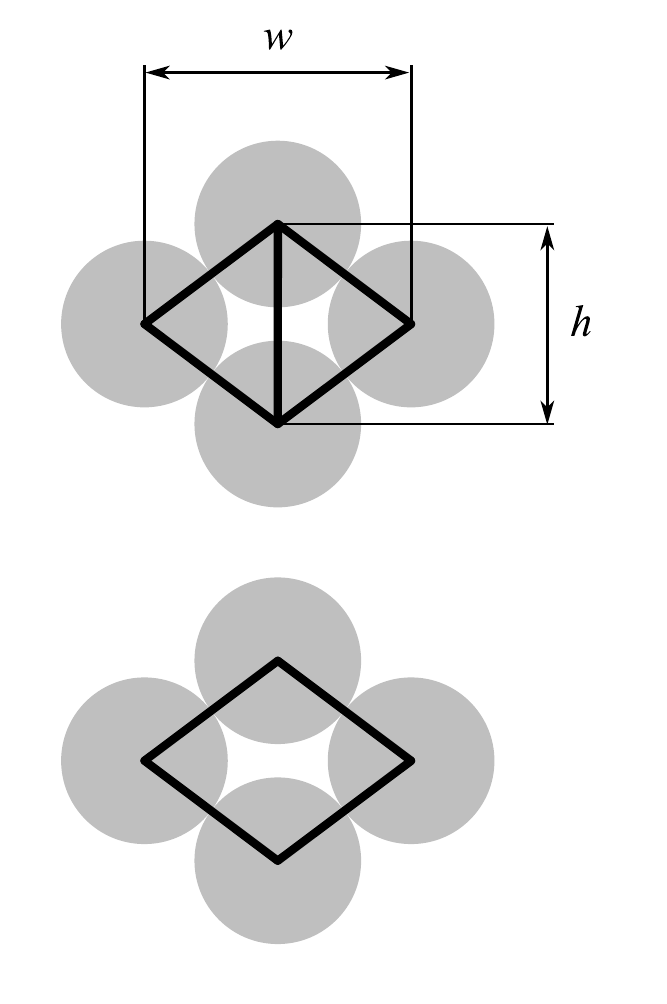}

\caption{Atoms arranged on the corners of a rhombus with height $h$ and width
$w$. The atoms could be identified as forming two SP3 rings as shown
above or as forming a single SP4 ring as shown below. Which of these
possibilities is identified will depend on the choice of $f_{c}$
appearing in Eq. \ref{eq:mod-voronoi}. \label{fig-2}}
\end{figure}

To demonstrate how our modified Voronoi criteria can effectively identify
SP4 rings we consider four atoms which are all located in the same
plane, forming a rhombus, Fig \ref{fig-2}. %
 The rhombus is characterised by the ratio $h/w$. Using the standard
Voronoi method, $f_{c}=1$, the only rhombus that will be identified
as forming an SP4 ring is a square $h/w=1$. As we reduce the value
of $f_{c}$ from unity, more asymmetric rhombuses are identified as
forming an SP4 ring. If $f_{c}=0.9$ the most asymmetric rhombus which
will be identified as forming an SP4 ring is $h/w=0.9045$ and the
maximum radius for bonding to atom $i$ as depicted in Fig. \ref{fig-1}
with the minimum separation distance set to $r_{st}=0.9$ is $r_{c}\simeq3.791$.
For $f_{c}=0.8$ we have $h/w=0.8165$ and $r_{c}\simeq2.504$. The
minimum value of $h/w$ for which the disks depicted in Fig. \ref{fig-2}
will not overlap is $h/w=0.5774$ which would require a value of $f_{c}=0.5$
to be identified as forming an SP4 ring.

\section{Identifying the Clusters}

To identify the clusters we will assume that many of the possible
configurations can be discounted due to the steric interaction between
the atoms. The effective use of this assumption, in terms of what
may be discounted, has been refined by trial and error, both in terms
of identifying the various clusters in isolation and by applying the
code to actual data and checking the results using molecular viewing
software. The strategies used to identify the clusters can depend
on the method used to establish the bond network. We have successfully
employed two methods. First declaring atoms to be bonded when they
are very close to the minimum separation distance allowed by their
steric interaction. We have used this method to good effect in analysing
microscopy data obtained from colloidal gels \cite{Royall-tbp-07}.
These gels feature a potential interaction between the colloidal particles
which have a very narrow deep well, providing a natural bond length.
Secondly we have used the modified Voronoi method with the parameter
$f_{c}=0.82$ (see Eq. \ref{eq:mod-voronoi}) and the cutoff radius
$r_{c}=2.0$ for the results reported in section VI here.

After identifying a cluster we store the indices of its atoms which
are used later in the final analysis. For this reason any redundant
identification of clusters will not concern us. The description below
makes no effort to describe how things may be computed efficiently.
Rather it attempts to be concise.

Below we describe the features of the various clusters which are used
to identify them. The reader may find it helpful to obtain the configuration
files for the various clusters \cite{morse-clusters}, and refer to
the way they are formed from the various basic clusters using a molecular
viewing program. A few example images are provided in Fig \ref{fig:more-clusts}.

\subsection*{The Basic Clusters}

\begin{figure}[H]
\includegraphics[scale=0.3]{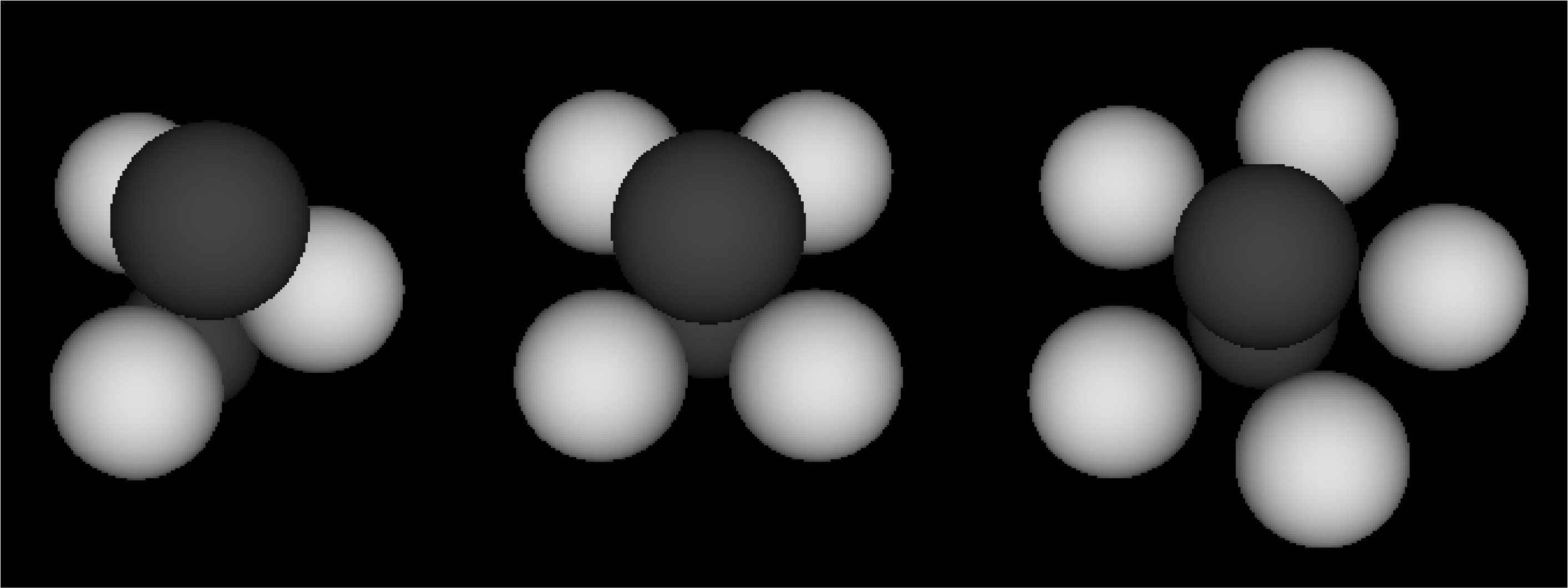}

\caption{The SPnc clusters. The spindle atoms are dark and the ring atoms
are light. From left to right we have SP3c/5A, SP4c/6A, SP5c/7A.\label{fig:basic-c}}
\end{figure}

After obtaining the bond network we identify all the SP3, SP4 and
SP5 rings in the system. A method by which this may be done is given
in reference \cite{FRANZBLAU1991}. We then divide each of the rings,
of a given size, into a further three types. For the first of these
there are an additional two atoms which are bonded to each atom in
the ring. We assume that the steric interaction is such, that the
only way this can occur is for one atom to be bonded on each side
of the ring (see Fig. \ref{fig:basic-c}), and that it is not possible
to have more than two atoms bonded to all the atoms forming one of
these rings. We will denote such clusters as SP\emph{n}c where \emph{n}
is the number of atoms in the SP ring. These SP\emph{n}c clusters
correspond to the first three Morse clusters, which we have now succeeded
in identifying. Using the notation given in \cite{DOYE1995,morse-clusters}
we have SP3c = 5A, SP4c = 6A and SP5c = 7A. The next type of basic
cluster we identify is a ring that has only one additional atom bonded
to all its atoms. We denote such a cluster as SP\emph{n}b. Note that
the 4 atoms forming an SP3b cluster form a total of 4 SP3b clusters.
This redundancy will not concern us. All rings which have no additional
atoms bonded to all of their atoms are labelled as an SP\emph{n}a
cluster. 

The additional two atoms forming an SP\emph{n}c cluster will be referred
to as the spindle atoms. Under the definitions given here, it is possible
that these two spindle atoms are or are not bonded to each other.
The single additional atom of the SP\emph{n}b clusters is also referred
to as a spindle atom. 

One could take an SPnc cluster and find it contains SPnb and SPna
clusters. By construction we have imposed the restriction that a given
SPn ring is used to form one cluster only, be it type a, b or c. For
the larger clusters, that we deal with next, no restriction will be
made about whether they are contained in other clusters or not. %

\subsection*{The Eight and Nine Membered Clusters}

\begin{figure}[h]
\includegraphics[scale=0.5]{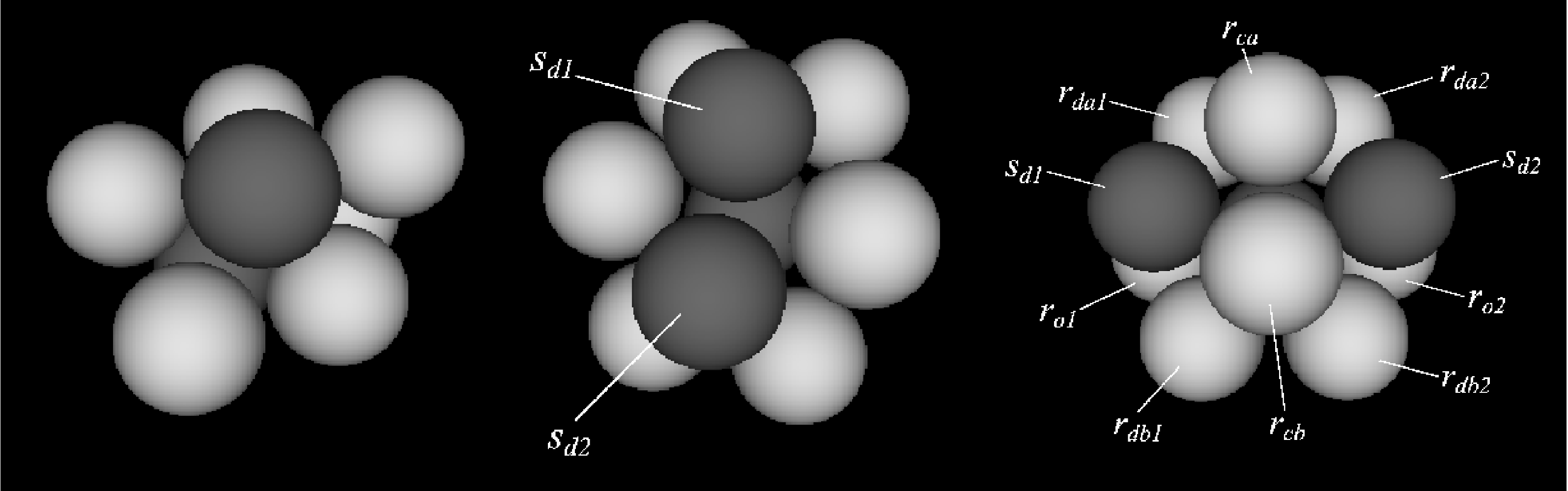}

\caption{On the left is the 8A cluster with the spindle atoms of the SP5c
or SP5b clusters coloured dark. In the middle is the 9B cluster with
the two distinct spindle atoms labelled and the common spindle atom
$s_{com}$ (coloured dark) partially visible around the back. On the
right is the 11C/11D cluster with all of the atoms labelled except
for the common spindle atom $s_{com}$ which is the dark one partially
visible, centre back.\label{fig:more-clusts}}
\end{figure}

The first of the eight membered clusters is the 8A Morse cluster,
see Fig. \ref{fig:more-clusts}. We chose to identify this in terms
of SP5 rings. It is also possible to identify it in terms of SP4 rings
which could be a good choice when using the modified Voronoi method,
depending on the value chosen for $f_{c}$. If desired, one could
do both. The method we use here will sometimes identify the same structure
twice, but we are not concerned with this as we can correct for it
later. We first search through all possible pairs of SP5b clusters.
If we find a pair which has different spindle atoms, $s_{1}$ \& $s_{2}$,
and 4 atoms which are common to both the SP5 rings we label the 8
atoms as forming an 8A cluster. We then search all pairs of SP5c clusters
and any pair which has in common the two spindle atoms and 4 SP5 ring
atoms, with the 5th ring atoms distinct (i.e. a member of only one
of the pair of basic clusters under consideration), is labelled as
forming an 8A cluster. Finally we search all pairs composed of one
SP5b and one SP5c cluster. If the SP5b spindle atom is also a spindle
atom for the SP5c cluster and there are 4 common SP5 ring atoms we
label it an 8A cluster. %

The 8B cluster is composed of an SP5c cluster and one additional atom.
For every SP5c cluster we search through all the other atoms and each
time we find one which is bonded to two of the SP5 ring atoms and
one spindle atom we have found an 8B cluster.

The 9A cluster is formed from a triplet of SP4b clusters. All combinations
of three SP4b clusters are searched to find those which have shared
atoms and bonds between the atoms from the three SP4 rings consistent
with the 9A cluster. In addition it is checked that the spindle atoms
are distinct and not bonded to each other.

The 9B cluster is formed from a pair of SP5c clusters, Fig. \ref{fig:more-clusts}.
The pair has one spindle atom which is common to both basic clusters,
$s_{com},$ and the other two spindle atoms are distinct and bonded
to each other ($s_{d1},s_{d2}$). The two distinct spindle atoms are
members of the other basic cluster's SP5 ring. The two SP5 rings have
two atoms in common. Each pair of SP5c rings are checked for these
conditions.

\subsection*{The Ten and Eleven Membered Clusters}

The 10A cluster is formed from a pair of SP4b clusters. None of the
atoms are common and the spindle atoms are not bonded to any of the
other basic cluster's atoms. Each atom of one SP4 ring is bonded to
two atoms of the other SP4 ring.

The 10B cluster is composed of three SP5c clusters. It is also composed
of three 9B clusters. We start with an 9B cluster and search all SP5c
clusters which have an index higher than both the SP5c clusters forming
the 9B cluster. This ensures we don't search the same combinations
more than once. One of the new SP5c cluster's spindle atoms is common
to the common spindle atom in the 9B cluster, $s_{com}$. The other
spindle atom, $s_{d3}$, is bonded to both the distinct spindle atoms
of the 9B cluster forming the SP3 ring ($s_{d1},s_{d2},s_{d3}$).

The 11A cluster contains SP4 rings with bond lengths which are not
particularly close together. For this reason it is not well suited
to identification by the bond length method, although it can work,
the Voronoi method will be superior for this case. Pairs of SP4c clusters
are searched. The clusters have a common spindle atom, $s_{com}$,
with all other atoms being distinct. Each of the SP4 ring atoms is
bonded to 2 atoms from the other SP4 ring.

The 11B cluster is formed from an 9B cluster and two additional atoms,
($e_{1}$, $e_{2}$). These two additional atoms are bonded to the
common spindle atom of the 9B cluster, $s_{com}$, and bonded to each
other. The two extra atoms also form four bonds with four different
atoms from the two SP5 rings. These four SP5 atoms form two pairs
which are bonded to each other with no bonds across the pairs. 

The 11C and 11D clusters are equivalent for our purposes. They can
be formed from two SP5c clusters with one common spindle atom $s_{com}$.
The other two distinct spindle atoms are not bonded, $s_{d1}$ \&
$s_{d2}$. The two SP5 rings have two common atoms which are bonded
to each other ($r_{ca},r_{cb}$). There are two bonds, between the
distinct atoms from the two different SP5 rings, involving four atoms,
($r_{da1},r_{da2}$) \& ($r_{db1},r_{db2}$). See Fig. \ref{fig:more-clusts}.

The 11E cluster looks somewhat different to the clusters we have considered
so far. None the less it can be constructed by combining an 9B cluster
with an additional SP5c cluster. One of the spindle atoms of the SP5c
cluster, $s_{com2}$, is common with one of the distinct spindle atoms
from the 9B cluster, $s_{com2}=s_{d1}$. The other spindle atom, $s_{d3}$,
of the new SP5c cluster is bonded to the other distinct spindle atom,
$s_{d2}$, and also bonded to the common spindle atom, $s_{com}$,
of the 9B cluster.

The 11F cluster can be formed from a combination of two SP3c and two
SP4c clusters. To identify it we first search through all pairs of
SP3c clusters. The spindle atoms of the SP3c clusters are distinct
and are all bonded to one of the spindle atoms from the other cluster.
Thus we have two bonded pairs of spindle atoms ($s_{t1},s_{t2}$)
\& $(s_{b1},s_{b2})$. The SP3 rings have one common atom, $r_{com}$,
and one bonded pair of distinct atoms from the different SP3 rings
($r_{c1},r_{c2}$). The SP4c clusters have $r_{com}$ as one of their
spindle atoms with the other spindle atom being new, $s_{e1}$ and
$s_{e2}$. The SP4 ring of the first SP4c cluster consists of ($r_{c1},r_{c2},s_{t1},s_{t2}$)
and that of the second cluster consists of ($r_{c1},r_{c2},s_{b1},s_{b2}$).

\subsection*{The Twelve and Thirteen Membered Morse Clusters}

The 12A cluster may be formed from an 11C cluster with one additional
atom. The two rings, from the SP5c clusters we used to form the 11C
cluster, each have one atom that is not bonded to either of the rings
common atoms ($r_{ca},r_{cb}$): we label this pair of atoms, $r_{o1},\, r_{o2}$,
see Fig \ref{fig:more-clusts}. The extra atom is only bonded to three
atoms from the 12A cluster which are $s_{com},r_{o1},r_{o2}$.

The 12B and 12C clusters are the same for our purpose. They are essentially
an icosahedral cluster with one atom missing. This can be formed from
six SP5c clusters. There will be one central SP5c cluster with one
common spindle atom $s_{com}$ and one distinct $s_{dis}$. If we
can find an additional five SP5c clusters which have one spindle atom
given by $s_{com}$ and the other spindle atom bonded to $s_{dis}$
we have a 12B/12C cluster.

The 12D cluster is formed from an 11E cluster combined with an additional
SP5c cluster. The new SP5c cluster has one spindle atom which is common
with $s_{d3}$ and the other is common with $s_{d2}$ of the 11E cluster.

The 12E cluster is formed from an 11F cluster combined with an SP3c
cluster. The SP3c cluster has the $s_{e1}$ and $s_{e2}$ atoms, of
the 11F cluster, as its spindle atoms 

The 13A cluster is an icosahedral cluster. It can be formed from an
12B/12C cluster combined with an additional SP5c cluster. One of the
additional SP5c's spindle atoms is $s_{com}$. The other spindle atom
and all of the SP5 ring atoms are distinct from the central SP5c cluster
of the 12B/C cluster.

The 13B cluster is formed from two SP5c clusters. There is one common
spindle atom, $s_{com}$, with the other spindle atoms being distinct
and not bonded to each other. Every atom from the SP5 ring of the
first cluster is bonded to exactly one atom from the SP5 ring of the
second cluster.

\subsection*{The FCC and HCP Crystal Clusters}

The HCP cluster is formed from three SP3c clusters. The three SP3
rings have one atom in common, $r_{com}$, which is the only atom
any of the SP3c clusters have in common. The spindle atoms form two
SP3 rings, ($s_{a1},s_{a2},s_{a3}$) \& ($s_{b1},s_{b2},s_{b3}$).
Apart from the common atom, $r_{com}$, the spindle atoms are not
bonded to any of the rings atoms from the other SP3c clusters. Upon
excluding, $r_{com}$, from the SP3 rings of the three SP3c clusters
we are left with six atoms which form a six membered ring. This ring
is not a shortest path ring.

The FCC cluster can be formed from four SP3b clusters or from three
SP3b clusters and one SP3c cluster. The first three SP3b clusters
all have a common SP3 ring atom, $r_{com}$, and the spindle atoms
are all distinct forming an SP3 ring, ($s_{a1},s_{a2},s_{a3}$). Excluding
$r_{com}$ we are left with six atoms from the SP3b clusters SP3 rings,
which form a six membered ring. Again the six membered ring is not
a shortest path ring and when we combine it with atom $r_{com}$ we
obtain six SP3 rings. Three of these rings are from the initial SP3b
clusters and three are not. We will refer to the later as the three
new SP3 rings. The forth SP3b cluster or the SP3c cluster has $r_{com}$
as a spindle atom. Each of its SP3 ring atoms can be combined with
a different new SP3 rings to form an SP3b cluster. If an SP3c cluster
is used the second spindle atom is not part of the FCC cluster.

\section{Analysing The Results}

It is a simple matter to record the atoms which form the various clusters.
If some of the clusters have been identified multiple times this can
be checked for and corrected later. 

However the reporting of population levels for the various clusters
opens up choice and ambiguity. This is because a given atom may be
a member of several different clusters. We have decided to report
the population levels in the following manner. If an atom is a member
of a cluster and also a member of a different cluster which has more
atoms it is only identified with the larger cluster. An atom may be
a member of two clusters consisting of the same number of atoms, in
this case the atom is reported as being a member of both clusters
if it is not a member of any larger clusters. Using this approach
we can construct a histogram of the net population levels for the
various clusters.

\section{An Illustration: The Hard Sphere Fluid}

Hard spheres are seen as a basic reference model for the liquid state
whose structure is determined by the short range repulsive interaction
between the constituent atoms. This idea can be traced back to van
der Waals and was later developed, in the form of perturbation theory,
to the point where it could quantitatively account for the properties
of real simple liquids \cite{Barker-RMP-1976}.%
\begin{figure}[h]
\includegraphics[scale=0.5]{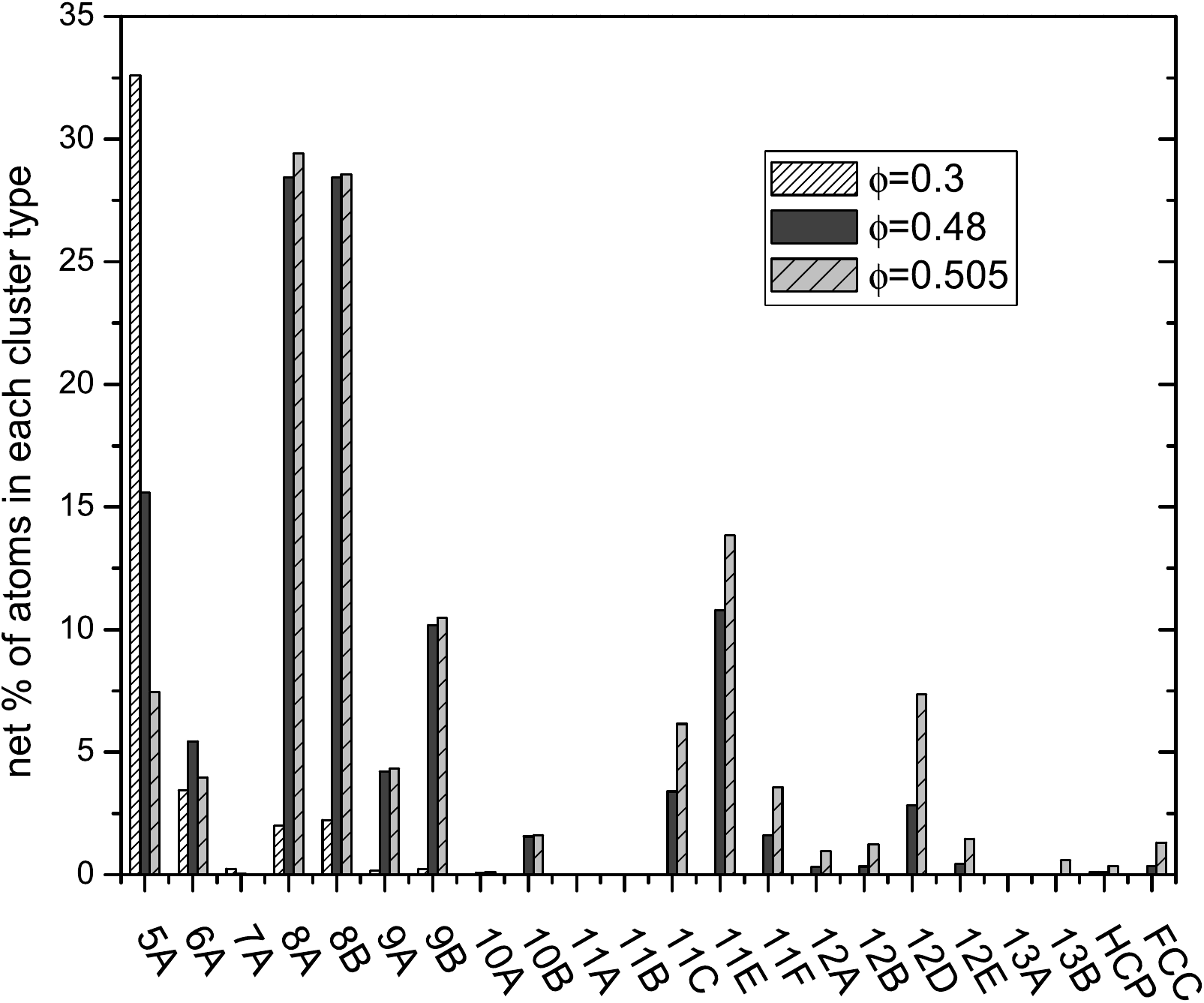}\\

\caption{A histogram of the net population levels for the various clusters
as discussed in Section V.\label{fig-histogram}}
\end{figure}
\begin{figure}[h]
\includegraphics[scale=0.3]{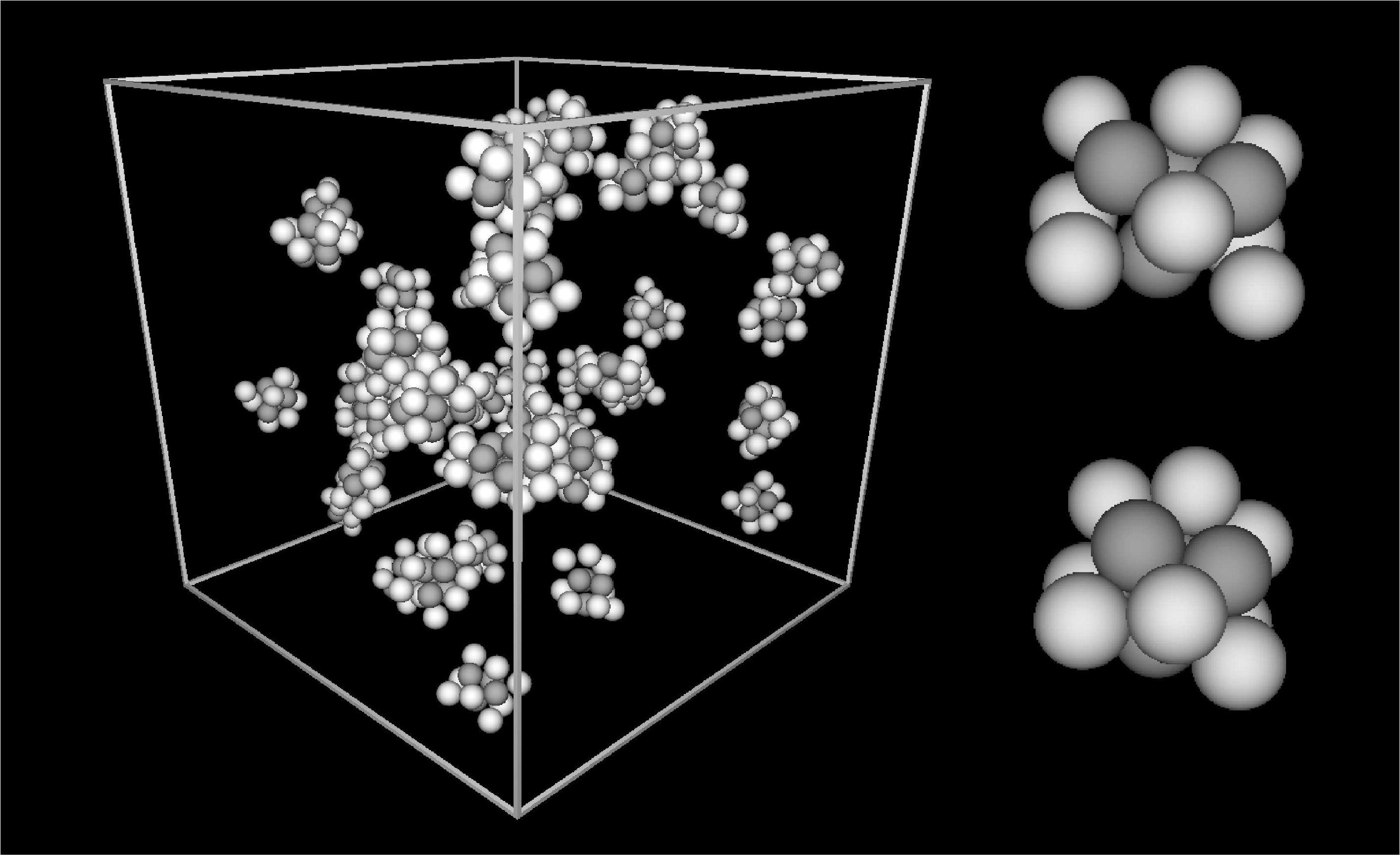}

\caption{On the left: the atoms which have been identified as members of a
12D cluster from the $\phi=0.505$ configuration. Four SP5c clusters
can be found in a 12D cluster, the spindle atoms from these SP5c clusters
are the darker coloured atoms. If an atom is a spindle cluster in
one 12D cluster and not in another it is still labelled with the dark
colour. Top right is a close up of one of the clusters from the configuration
and the bottom right is the configuration of the 12D Morse cluster
obtained from \cite{morse-clusters}. \label{fig-12D}}
\end{figure}
 At high densities (volume fractions) the efficient packing of the
crystal phase results in more entropy, relative to the disordered
fluid phase, for hard spheres and thus there is a first order phase
transition. The hard sphere fluid freezes at a volume fraction of
$\phi_{f}=0.494$ and melts at $\phi_{m}=0.548$ \cite{Hoover-JCP-1968}.
At low volume fractions, $\phi<\phi_{f}$, the fluid is the equilibrium
phase, between the freezing and melting volume fractions there is
a fluid at $\phi_{f}$ coexisting with an FCC crystal at $\phi_{m}$,
and at high volume fractions, $\phi>\phi_{m}$ the FCC crystal is
the equilibrium phase \cite{Bolhuis-N-1997}. 

We decided to test the new scheme on the hard sphere fluid phase at
volume fractions of $\phi=0.3,\,0.48\,\&\,0.505$. All simulations
used 10,976 atoms and a single configuration was analysed in each
case. The highest volume fraction is only marginally above the freezing
volume fraction and is not able to crystallise on the simulation time
scale leaving us with a metastable fluid phase. A Histogram showing
the net population levels of atoms in each cluster, for the three
volume fractions, is shown in Fig. \ref{fig-histogram}. At the low
volume fraction of $\phi=0.3$, 41\% of the atoms are identified as
being members of at least one of the clusters. The vast majority of
these are found to be members of a 5A (SP3c) cluster only. At the
higher volume fraction of $\phi=0.505$ things have changed considerably.
Here 99\% of the atoms are identified as being members of at least
one of the clusters. There is also a significant amount of larger
clusters being formed with the 8A and 8B cluster being most numerous.
There are very few 7A (SP5c) clusters in the histogram, apparently
this cluster is very likely to be part of a larger cluster. Perhaps
most striking is the significant number of 11E and 12D clusters which
are found. It has recently been speculated that the formation of clusters
around freezing could be responsible for the anomalous tail in the
velocity autocorrelation function \cite{Williams-PRL-2006,Hofling-PRL-2007}.
Whether this work can shed new light on this problem remains to be
seen.

Because the 12D clusters from the $\phi=0.505$ configuration are
the largest clusters identified in significant numbers and because
they are somewhat different to the icosahedral clusters that might
be expected \cite{FRANK1952} we focus on these. In Fig. \ref{fig-12D}
an image of the atoms which have been identified as members of 12D
clusters is shown. A close up of one of these clusters is also shown.
This cluster was chosen from the configuration at random. For comparison
an image of the actual Morse cluster configuration, taken from ref.
\cite{morse-clusters}, is also shown. The similarity, in the way
the atoms are arranged in both cases, can be readily recognised from
these images. Experience of inspecting the various clusters suggests
that this is regularly the case.

\section{Conclusions}

A new methodology to identify local structure in dense phases has
been introduced. This shows how studies on small isolated clusters
of atoms can provide important new insight into our understanding
of bulk phases. Because a given atom can be a member of several different
clusters the clusters may overlap and fill space. Remarkably some
99\% of the atoms in a hard sphere fluid, marginally above the freezing
volume fraction, are identified as being a member of at least one
of the Morse clusters. As the volume fraction is increased, still
more atoms will be identified as cluster members. This demonstrates
the promise of the approach, introduced here, as a powerful new tool
to further our understanding of bulk phases. In particular one would
expect important findings about supercooled liquids, vitrification
and possibly crystallisation to be made in the future using this methodology.

\begin{acknowledgments}
I thank C. (Paddy) Royall for many helpful discussions and for rekindling
my interest in this problem. I thank Denis J. Evans for support and
encouragement. 
\end{acknowledgments}
\bibliographystyle{apsrev}
\bibliography{clust2}
 
\end{document}